\documentclass[article,superscriptaddress,twocolumn]{revtex4-1}
\usepackage{graphicx}
\usepackage{dcolumn}
\usepackage{inputenc}
\usepackage{bm}
\usepackage{color}
\usepackage{graphicx}
\usepackage[english]{babel}
\usepackage{color}
\usepackage{booktabs,longtable}
\usepackage{natbib}
\usepackage{xspace}
\usepackage{textgreek} 
\usepackage{upgreek} 
\usepackage{mathrsfs}
\usepackage[colorlinks]{hyperref}
\hypersetup{
                pdfstartview={FitH},
                linkcolor=blue,
                citecolor=blue,
                filecolor=blue,
                urlcolor=blue
}

\newcommand{\fig}[2]{\textcolor{blue}{Fig.~\ref{Fig#1} #2)}}
\newcommand{\figsimple}[1]{\textcolor{blue}{Fig.~\ref{Fig#1}}}

\newcommand{\degree}{$^{\circ}$\xspace}
\newcommand{\kelv}{$~\textup{K}$\xspace}

\newcommand{\uA}{$~\text{\textmu A}$\xspace}

\newcommand{\BST}{(Bi$_{0.4}$Sb$_{0.6}$)$_2$Te$_3$\xspace}
\newcommand{\FGT}{Fe$_3$GeTe$_2$\xspace}
\newcommand{\BaF}{BaF$_2$\xspace}
\newcommand{\Rahe}{$~R_{\textup{AHE}}$\xspace}
\newcommand{\Rphe}{$~R_{\textup{PHE}}$\xspace}

\newcommand{\Rxy}[1]{$~R_{xy}^{#1\omega}$\xspace}
\newcommand{\BDL}{$~B_{\textup{DL}}$\xspace}
\newcommand{\BFL}{$~B_{\textup{FL}}$\xspace}

\begin{document}

\begin{sloppypar}

\title{Spin-orbit torque switching in 2D ferromagnet / topological insulator heterostructure grown by molecular beam epitaxy}

\author{T. Guillet}
\affiliation{Catalan Institute of Nanoscience and Nanotechnology (ICN2), CSIC and the Barcelona Institute of Science and Technology (BIST), Bellaterra, Barcelona, Spain}
\author{R. Galceran}
\affiliation{Catalan Institute of Nanoscience and Nanotechnology (ICN2), CSIC and the Barcelona Institute of Science and Technology (BIST), Bellaterra, Barcelona, Spain}
\author{J. F. Sierra}
\affiliation{Catalan Institute of Nanoscience and Nanotechnology (ICN2), CSIC and the Barcelona Institute of Science and Technology (BIST), Bellaterra, Barcelona, Spain}
\author{M. V. Costache}
\affiliation{Catalan Institute of Nanoscience and Nanotechnology (ICN2), CSIC and the Barcelona Institute of Science and Technology (BIST), Bellaterra, Barcelona, Spain}
\author{M. Jamet}
\affiliation{Univ. Grenoble Alpes, CEA, CNRS, Grenoble INP, IRIG-SPINTEC, 38000 Grenoble, France}
\author{F. Bonell}
\affiliation{Univ. Grenoble Alpes, CEA, CNRS, Grenoble INP, IRIG-SPINTEC, 38000 Grenoble, France}
\author{S. O. Valenzuela}
\affiliation{Catalan Institute of Nanoscience and Nanotechnology (ICN2), CSIC and the Barcelona Institute of Science and Technology (BIST), Bellaterra, Barcelona, Spain}
\affiliation{Instituci\'{o} Catalana de Recerca i Estudis Avan\c{c}ats (ICREA)
Barcelona 08010, Spain}

\date{\today}

\begin{abstract}

Topological insulators (TIs) are a promising class of materials for manipulating the magnetization of an adjacent ferromagnet (FM) through the spin-orbit torque (SOT) mechanism. However, current studies combining TIs with conventional FMs present large device-to-device variations, resulting in a broad distribution of SOT magnitudes. It has been identified that the interfacial quality between the TI and the FM is of utmost importance in determining the nature and efficiency of the SOT. To optimize the SOT magnitude and enable ultra-low-power magnetization switching, an atomically smooth interface is necessary. To this end, we have developed the growth of a full van der Waals FM/TI heterostructure by molecular beam epitaxy. The compensated TI \BST and ferromagnetic \FGT (FGT) were chosen because of their exceptional crystalline quality, low carrier concentration in BST and relatively large Curie temperature and perpendicular magnetic anisotropy in FGT. We characterized the magnitude of the SOTs by using thorough harmonic magnetotransport measurements and showed that the magnetization of an ultrathin FGT film could be switched with a current density J$_c$ $\leq$ 10$^{10}$ A/m$^2$. In comparison to previous studies utilizing traditional FMs, our findings are highly reliable, displaying little to no variation between devices.

\end{abstract}

\maketitle


One of the main challenges being faced by modern electronics is the need to reduce power consumption in advanced technology nodes, including both data processing and storage. To help tackle this challenge, magnetic random access memories have evolved through three generations of writing mechanisms, starting from the application of an Oersted field, with large power requirements and poor scalability, to spin-transfer torque and, more recently, to spin-orbit torques (SOT), which is currently under development\cite{bhatti_spintronics_2017}. The SOT mechanism utilizes the spin-orbit interaction to generate a spin current, providing several key advantages, such as increased writing speed, enhanced device durability, and reduced energy consumption \cite{manchon_current-induced_2019}. In the first SOT demonstrations, heavy metals (HM) such as Pt or Ta were used to induce the spin current and manipulate the magnetization of a conventional ferromagnetic (FM) material such as Co or CoFeB \cite{miron_current-driven_2010,liu_spin-torque_2012}. The strength of the SOT is primarily determined by the spin Hall angle of the HM, with the Rashba-Edelstein effect (REE) at the HM/FM interface likely contributing as well \cite{manchon_current-induced_2019}.

An emerging approach to enhance the SOT relies on replacing the traditional HMs by novel materials such as topological insulators \cite{mahendra_room-temperature_2018,wang_current-driven_2019} (TIs) or Weyl semimetals \cite{fan_quantifying_2014,macneill_control_2017,li_spin-momentum_2018,zhao_unconventional_2020}. Here, topological surface states (TSS) and the Rashba interface states present the potential to generate record SOTs.
However, results obtained to date with TIs have exhibited significant dispersion \cite{wang_topological_2015,shi_efficient_2018}. The poor interface sharpness and the high chemical affinity between Bi-based TIs and classical 3\emph{d} FMs are major hurdles preventing the achievement of the predicted breakthrough in magnetization switching power efficiency \cite{kondou_fermi-level-dependent_2016,bonell_control_2020}.

The discovery of two-dimensional van der Waals (vdW) FMs \cite{huang_layer-dependent_2017,may_ferromagnetism_2019,hu_recent_2020} could provide a solution to this issue \cite{fujimura_current-induced_2021,yang_two-dimensional_2022}. Several groups have observed robust ferromagnetism in \FGT (FGT) and related materials on both exfoliated flakes \cite{fei_two-dimensional_2018} and MBE-grown thin films \cite{wang_above_2020,lopes_large-area_2021,ribeiro_large-scale_2022}. The vdW nature of the interaction between the TI and a vdW FM can help limit unwanted chemical reactions, atomic intermixing and electronic hybridization at the interface.
In this work, we focus on vdW bilayers made of FGT and a high-quality compensated TI \BST (BST) grown by molecular beam epitaxy \cite{bonell_growth_2017}. We employ high magnetic field second harmonic magnetotransport measurements to characterize the SOT \cite{garello_symmetry_2013}, finding very large torques with strong temperature dependence. Additionally, we report current-induced magnetization switching requiring a very low critical current density (J$_c$ $\leq$ 10$^{10}$ A/m$^2$).

	
FGT/BST bilayers were grown epitaxially by MBE on \BaF(111) \cite{bonell_control_2020,bonell_growth_2017}. The vdW character of the layers, combined with their shared six-fold symmetry, leads to the alignment of the FGT and BST crystalline axes as evidenced in their corresponding RHEED patterns, despite the considerable lattice mismatch between the two crystals (4.7 $\%$).
Two samples were grown: 7 nm of FGT on 10 nm of BST (BST/FGT(7)) and 2 nm of FGT on 10 nm of BST (BST/FGT(2)).
These two samples were then patterned into Hall bars to perform in depth magnetotransport measurements.


First, we analyze the temperature-dependent longitudinal resistivity $\rho(T)$ and anomalous Hall effect (AHE) of the bilayer in order to gain insight into its magnetic properties and the current distribution. The geometry of the measurements is illustrated in \fig{1}{a}. Here, the current is applied along $x$ and the polar and azimuthal angles of the magnetic field $B$ and magnetization $M$ vectors are denoted as ($\theta_{B}$, $\varphi_{B}$) and ($\theta_{M}$, $\varphi_{M}$), respectively, using spherical coordinates. Because of the strong perpendicular magnetic anisotropy (PMA) of FGT, $\theta_{M}$ and $\theta_{B}$ are not equal at the applied $B$.

\begin{figure}[h!]
\begin{center}
\includegraphics[width=0.45\textwidth]{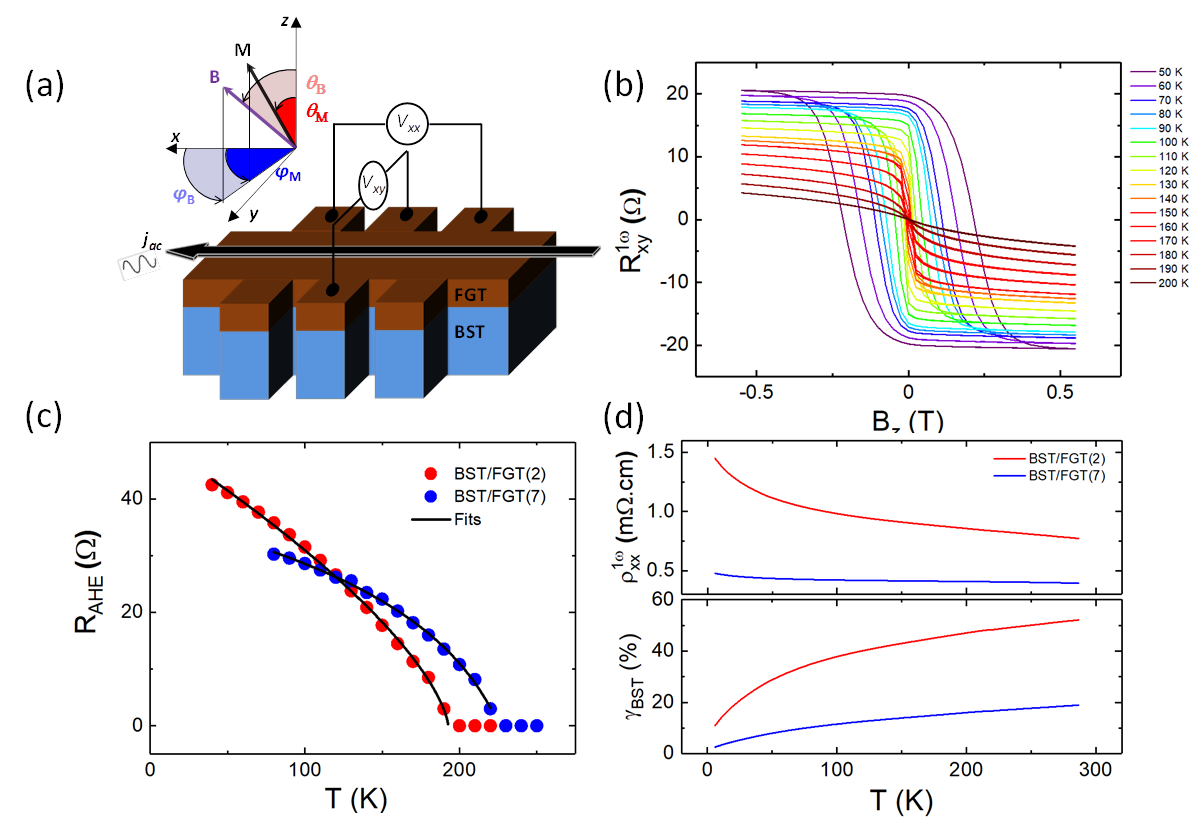}
\caption{(color online) a) 8-contacts Hall bar geometry of the devices used for the transport experiments. $\theta_{B}$ ($\theta_{M}$) and $\varphi_{B}$ ($\varphi_{M}$) denote the external magnetic field (FGT magnetization) polar and azimuthal angles using spherical coordinates, the current defines the $+x$ direction. b) AHE hysteresis loops of the BST/FGT(2) sample recorded at various temperatures with 10 \uA applied current. c) \Rahe temperature dependence indicating a thickness-dependent $T_{c}$. The black lines corresponds to fits according to the Curie-Weiss law. d) Top panel: Four-probe resistivity versus temperature for the two FGT thicknesses. Bottom panel: Corresponding calculated shunting coefficient in BST $\gamma_{BST} = \frac{\textup{J}_{\textup{BST}}}{\textup{J}_{\textup{BST}}+\textup{J}_{\textup{FGT}}}$. }
\label{Fig1}
\end{center}
\end{figure}

\fig{1}{b} shows AHE hysteresis loops recorded on BST/FGT(2) for temperatures ranging from 50\kelv to 200\kelv in 10 \kelv steps.
The large coercive field shows that the PMA remains strong despite the ultrathin character of the vdW film. \fig{1}{c} shows the temperature dependence of the saturation value of the AHE, \Rahe, for the two BST/FGT samples.
The magnitude of \Rahe is very large, as already reported \cite{kim_large_2018}. The Curie temperature $T_\mathrm{C}$ of BST/FGT(7) is 230 \kelv, which is comparable to the bulk value. However, for BST/FGT(2), $T_\mathrm{C}$ is observed to be lower, at 180\kelv.

The temperature dependence of the longitudinal resistivity $\rho(T)$ for the two samples (\fig{1}{d}) shows a semiconductor-like behavior, indicating that a significant amount of current is flowing through the BST film. However, in BST/FGT(7), $\rho(T)$ remains almost constant since more current flows through the metallic FGT.
To estimate the shunting ratio between BST and FGT, we solved a parallel resistor model, the result of which is shown in the bottom panel of \fig{1}{c}. We found that in BST/FGT(2nm) for temperatures above 40\kelv, the charge current is roughly equally distributed in the BST and FGT layers, which is optimal to generate large SOTs and to obtain sizeable second harmonic voltages.

Having characterized the transport and magnetic properties of the bilayer, we use the harmonic Hall method to extract the magnitude of the SOTs \cite{garello_symmetry_2013}. A low-frequency AC current (177 Hz) is injected in the Hall bar (along $x$), which generates a transverse spin current (along $z$) by the possible bulk spin Hall effect (SHE) as well as a spin accumulation at the BST/FGT interface through the REE. In both cases, the spin orientation is along $y$. The accumulation of spin-polarized electrons at the interface gives rise to a field-like (FL) SOT acting on the FGT magnetization.
The spin-polarized current reaching the FGT layer, typically by the SHE, generates a damping-like (DL) SOT.
As the current is AC, the orientation of the spin reverses over one period, resulting in oscillatory torques.

The effect of these two distinct SOTs on the FGT magnetization can be described with the presence of two effective magnetic fields, \BDL and \BFL, which are characterized by their symmetry with respect to the magnetization: \BDL$\sim \mathbf{M} \times \mathbf{y}$ and \BFL$\sim \mathbf{y}$. As a result of these two effective fields, the magnetization oscillates in- and out-of-plane, which in turn affects the AHE and PHE signals. Following Ref. [\citenum{garello_symmetry_2013}], the first and second harmonics of the resistance are given by the following relationships:

\begin{equation} \label{eq1}
R_{xy}^{1\omega} = R_{\textup{AHE}}\cos \theta_M + R_{\textup{PHE}}\sin2\varphi_M \sin^{2}\theta_M
\end{equation}

\begin{equation}\label{eq2}
R_{xy}^{2\omega,{DL(FL)}} = \frac{dR_{xy}^{1\omega}}{dB_{zx(zy)}} \frac{B^{DL(FL)}}{\sin(\theta_{B}-\theta_{0})}
\end{equation}

\noindent where \Rphe represents the resistance amplitude of planar Hall effect (PHE) and $B_{zx(zy)}$ is the external magnetic field, applied in the $zx$ ($zy$) plane. In our samples, the ratio between the PHE and the AHE is smaller than 0.1 \% for the full temperature range (see Supp. Mater). Thus, the PHE is negligible in comparison to the AHE, simplifying the analysis of the SOT measurements \cite{avci_fieldlike_2014}.

\figsimple{2} summarizes the SOT measurements on the BST/FGT(2) sample, with an applied current of a 1 mA ($j = 2\times 10^{10}$A/m$^2$). \fig{2}{a} illustrates the magnetic-field configuration that allows us to probe the torque. To determine the DL torque, $B$ is applied in the $zx$ plane ($\phi_B=0$), close to the in-plane orientation ($\theta_{B} = 85$\degree) to be able to to ensure uniform magnetization at sufficiently large $B$. \fig{2}{b} shows the first-harmonic response \Rxy1. In comparison to \fig{1}{b}, the AHE loops appear wider, reflecting the smaller $B_z$ component in the present configuration. At high $B$, the magnetization gradually tilts toward the $x$ direction and the AHE signal eventually vanishes.

The DL torque contributes an \textit{antisymmetric} component to the measured second harmonic signal \Rxy2 as a function of $B$. To isolate $R_{xy}^{2\omega,{DL}}$, we have therefore anti-symmetrized \Rxy2 (see \fig{2}{c}). The magnitude of \Rxy2 is observed to be strongly enhanced at low $B$. This is due to the presence of magnetic domains and domain wall motion induced by the SOTs (See Supp. Mater.). Beyond the coercive field, determined at each temperature from \fig{2}{b}, the magnetization saturates and coherently rotates as $B$ increases. In this region, the anti-symmetrized \Rxy2 scales linearly with $B^{zx}$, as expected for the DL torque (Eqs. \ref{eq1} and \ref{eq2}). Equation 2 suggests that \Rxy2 can be linearized as a function of $ X_{\textup{SOT}} = \frac{dR_{xy}^{1\omega}}{dB_{zx(zy)}} \frac{1}{\sin(\theta_{B}-\theta_{0})}$. The first derivative of \Rxy1 with respect to $B$ determines the AHE-related measurement sensitivity; $\theta_0$ (the polar angle of \textbf{M} at equilibrium) is obtained from $\theta_0 = \cos^{-1}(R_{xy}^{1\omega}/R_{\textup{AHE}})$.
The linearization is implemented in \fig{2}{e}. The observed linear response at every temperature validates our analysis; by fitting the slope, \BDL can be determined directly.

Note that the anomalous Nernst effect (ANE) resulting from an out-of-plane thermal gradient (ANE$_z$) can contribute a signal with the same symmetries as the DL torque. Furthermore, an hysteretic behavior due to the ANE could also be observed in the presence of an in-plane thermal gradient (ANE$_x$). Therefore, before extracting the DL torque, both ANE$_z$ and ANE$_x$ have been removed from the (high $B$) data (see Suppl. Mat.).

We repeat the previous approach for $B$ in the $zy$ plane, using the same polar angle. In this configuration, the FL torque contributes a \textit{symmetric} component to \Rxy2. The symmetrized \Rxy2 as a function of $B$ is shown in \fig{2}{d}. This symmetrization treatment allows us to remove spurious magnetothermal contributions, which are all antisymmetric with $B$, and isolate $R_{xy}^{2\omega,{FL}}$. As before, we focus on the high-$B$ response and implement the linearization analysis to extract \BFL (\fig{2}{f}).

\begin{figure}[h!]
\begin{center}
\includegraphics[width=0.45\textwidth]{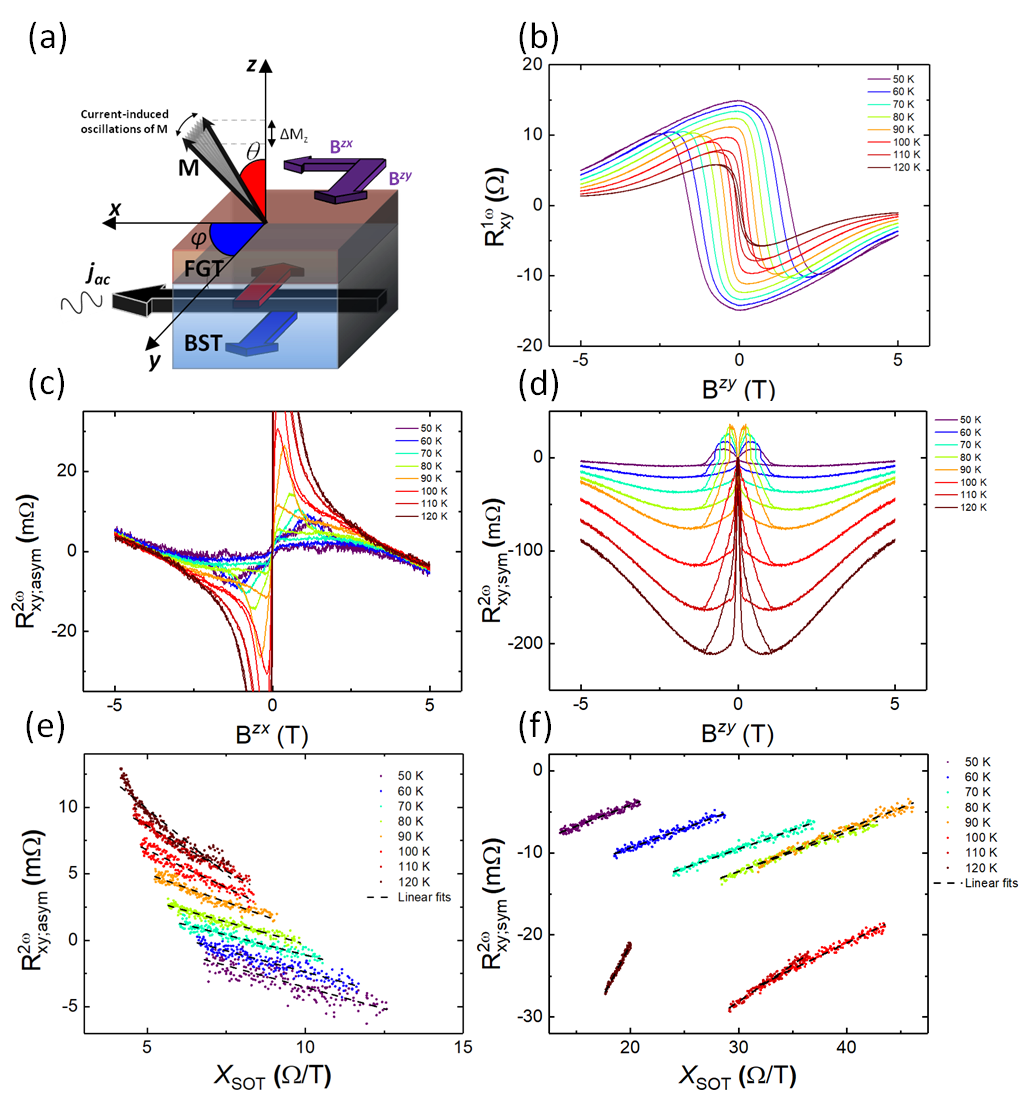}
\caption{(color online) a) Geometry of the SOT measurements, the current is applied along $+x$ and the external magnetic field is applied in the $zx$ ($zy$) plane, close to the in-plane magnetization orientation ($\theta_{B} = 85$\degree) to extract the damping-like (field-like) SOT. b) AHE loops for temperatures ranging from 50\kelv to 120\kelv , the high field reduction of \Rahe indicates the coherent in-plane tilting of the magnetization. c) and d) Antisymmetric (symmetric) part of \Rxy2 in the $zx$ ($zy$) space plane, characteristic of the DL (FL) SOT. e) and f) Corresponding linearized plots, obtained with the following variable change: $X_{\textup{SOT}} = \frac{dR_{xy}^{1\omega}}{dB_{zx(zy)}} \frac{1}{\sin(\theta_{B}-\theta_{0})}$.}
\label{Fig2}
\end{center}
\end{figure}

The magnitudes of \BDL and \BFL normalized by the current density in BST, $J_{BST}$ are shown in \figsimple{3}. Notably, the SOTs are particularly strong in the BST/FGT(2) bilayer and remain significant in BST/FGT(7), with a similar temperature dependence. As the temperature increases, the SOTs also increase, which can be attributed to the gradual vanishing of the FGT magnetization as the Curie temperature of FGT is approached.

\begin{figure}[h!]
\begin{center}
\includegraphics[width=0.45\textwidth]{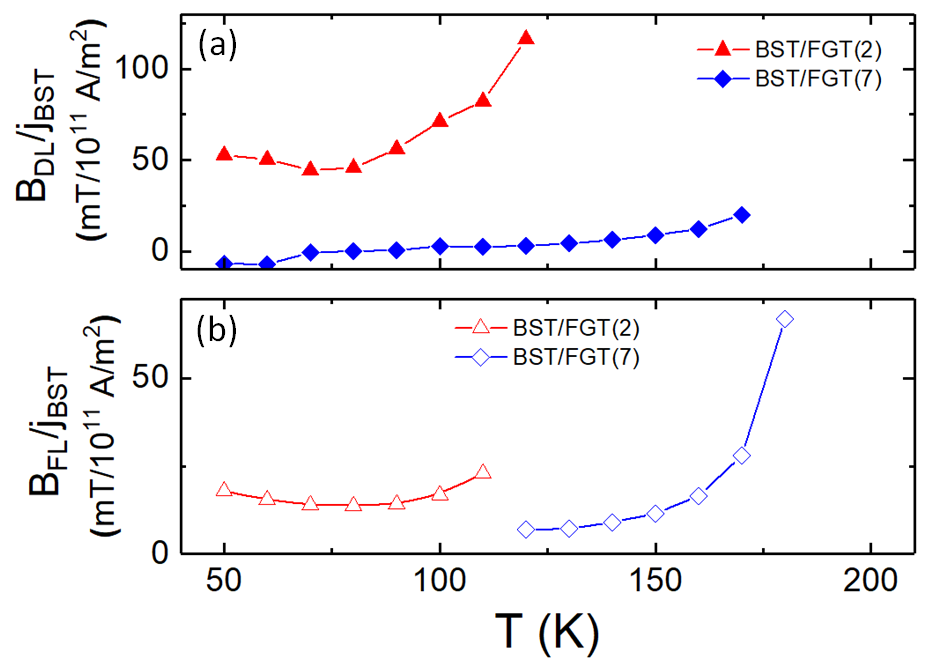}
\caption{(color online) Temperature dependence of the SOTs, BST/FGT(2) results are represented by red triangle (full for DL and opened for FL) and BST/FGT(7) data are shown as blue diamonds (full for DL and opened for FL).}
\label{Fig3}
\end{center}
\end{figure}


The large magnitude of the SOT effective fields suggests that current-induced magnetization switching should be achievable with relatively low current densities. To verify this assumption and determine the switching current, the magnetization is first initialized along $\pm z$, then a 200 mT assistance field along $\pm x$ is applied. The amplitude of 50 $\mu$s- current pulses is swept back and forth between $\pm 3.5$ mA and the AHE is recorded in DC conditions by applying a 10 \uA reading current. The assistance $B_x$ field is applied to facilitate the deterministic switching of the out-of-plane magnetization by the DL SOT.
The results of this experiment are summarized in \figsimple{4}. \fig{4}{a} shows the hysteresis loops obtained by sweeping a magnetic field up to 600 mT along $z$ at various temperatures from 50 to 125 K. These measurements serve as references. \fig{4}{b} shows the hysteresis loops obtained by sweeping the current pulse magnitude at the same temperatures. The change of chirality of the latter with the direction of the assistance field (represented with associated red and blue symbols) is a signature of SOT switching.

\begin{figure}[h!]
\begin{center}
\includegraphics[width=0.45\textwidth]{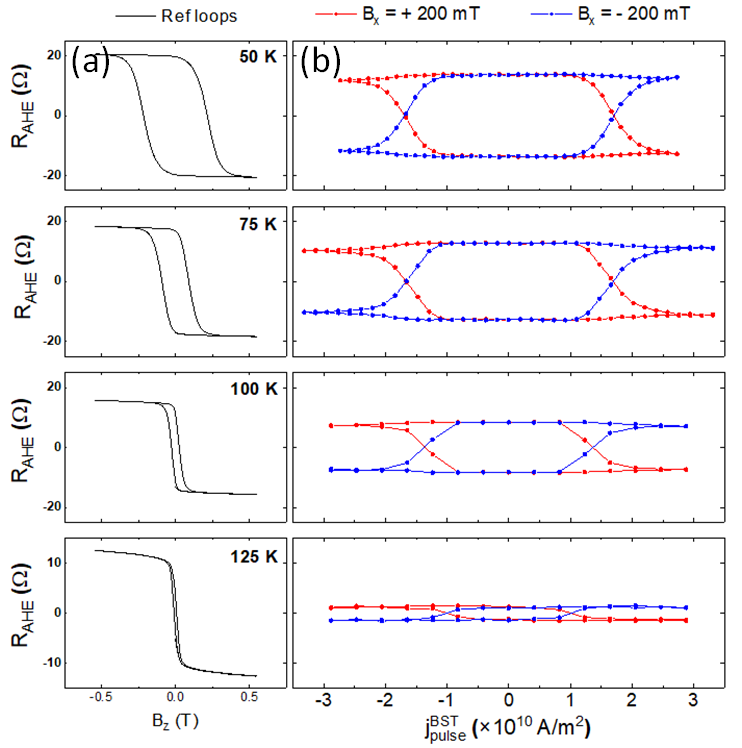}
\caption{(color online) SOT switching experiments a) AHE reference loops recorded by applying a perpendicular ($B_{z}$) magnetic field for temperatures ranging from 50\kelv to 125\kelv . b) Corresponding SOT switching hysteresis loops for positive and negative assistance magnetic field ($\pm B_{x} = 200$~mT).}
\label{Fig4}
\end{center}
\end{figure}

Notably, the switching current density $j_{\textup{sw}} \sim 1\times 10^{10}$A/m$^2$ is significantly smaller than in typical HM/FM heterostructures and it decreases with temperature as the magnetization of FGT decreases. However, by comparing the magnitude of the loops obtained by SOT switching to the reference ones, it is evident that the magnetization switches only partially. This is a consequence of the important Joule heating during the relatively long current pulse, which suppresses the FGT magnetization and the magnetic anisotropy of the FGT film (See Supp Mat.). These observations are promising and suggest that further optimization of vdW FM/TI heterostructures should allow us to reach even lower switching currents, while shorter current pulses should lead to full magnetization switching.

SOT switching of FGT has already been reported by a few groups. Utilizing Pt yielded $j_{\textup{sw}} = 2\times 10^{11}$A/m$^2$ \cite{alghamdi_highly_2019}. A full vdW heterostructure composed of WTe$_2$ and FGT fabricated with mechanically exfoliated FGT flakes \cite{shin_spinorbit_2022} led to a smaller $j_{\textup{sw}} = 4\times 10^{10}$A/m$^2$. A recent work reported the growth of BST/FGT by MBE\cite{fujimura_current-induced_2021}. The FGT layer was thicker than in the present work and resulted in a switching current from $0.4$ to $1 \times 10^{11}$A/m$^2$ at 100\kelv. Our fully epitaxial system exhibits exceptionally low switching current density $\sim 10^{10}$A/m$^2$ at 100\kelv. The large SOTs combined with the large PMA of FGT underline the interest of this vdW heterostructure. The temperature dependence of the DL and FL torques do not point towards a high contribution from the TSS.

In conclusion, we have developed the growth of a fully epitaxial vdW FM/TI heterostructure by molecular beam epitaxy. We demonstrated robust ferromagnetism in FGT down to 2 monolayers. Thanks to harmonic magnetotransport measurements, we found very strong spin-orbit torques generated by BST acting on the magnetization of FGT. We achieved FGT magnetization switching by using short current pulses with current densities $\sim 10^{10}$A/m$^2$, which represents a significant improvement compared to regular HM/FM heterostructures. Future studies are required to optimize the Sb composition and the BST thickness in order to maximize the contribution of the TSSs for ultrathin thin FGT films. Tuning the Curie temperature of FGT is an interesting prospect to develop a scalable ultra-low power magnetic memory.
	

This research has received funding from the European Union’s Horizon 2020 (EU H2020) research and innovation programme under grant agreement 881603 (Graphene Flagship). ICN2 acknowledges support from the Spanish Ministry of Science and Innovation (MCIN) and Spanish Research Agency (AEI/10.13039/501100011033) under contracts PID2019-111773RB-I00/, PCI2021-122035-2A, and Severo Ochoa CEX2021-001214-S. TG and RG acknowledge support from EU H2020 programme under the Marie Skłodowska-Curie grant agreements No. 754510 and No. 840588 (GRISOTO, Marie Sklodowska-Curie fellowship) respectively, and JFS from MCIN/AEI/10.13039/50110001103 under contract RYC2019-028368-I.

\end{sloppypar}

\bibliographystyle{unsrt}

\end{document}